\documentclass[prl,aps,twocolumn,showpacs,floatfix,superscriptaddress]{revtex4}
\usepackage{graphicx}
\usepackage{amsmath}
\usepackage{amssymb}

\begin{document}

\title{Equilibration of a spinless Luttinger liquid}

\author{K. A. Matveev} 

\affiliation{Materials Science Division, Argonne National Laboratory,
  Argonne, IL 60439, USA}

\author{A. V. Andreev} 

\affiliation{Department of Physics, University of Washington, Seattle, WA
  98195, USA}

\date{July 20, 2011}

\begin{abstract}

  We study how a Luttinger liquid of spinless particles in one dimension
  approaches thermal equilibrium.  Full equilibration requires processes of
  backscattering of excitations which occur at energies of order of the
  bandwidth.  Such processes are not accounted for by the Luttinger liquid
  theory.  We treat the high-energy excitations as mobile impurities and
  derive an expression for the equilibration rate in terms of their spectrum.
  Our results apply at any interaction strength.

\end{abstract}


\maketitle

The concept of Luttinger liquid was proposed by Haldane as an effective
low-energy description of one-dimensional systems of interacting fermions
\cite{haldane} or bosons \cite{haldane2}.  The main feature of this theory is
that regardless of the statistics of the particles, the low-energy excitations
of the system are bosons.  The latter propagate at a fixed velocity $v$ in
either left or right direction and have the meaning of the waves of particle
density, analogous to phonons in solids.

In its simplest form the Luttinger liquid is described by a Hamiltonian
quadratic in boson variables, resulting in excitations with infinite life
time.  Once excited, such a system will never reach thermal equilibrium.
Absence of equilibration is the physical reason \cite{conductance} for perfect
quantization of conductance of a quantum wire connected to ideal leads, when
the electronic system in the wire is treated as a Luttinger liquid
\cite{maslov}.  Of course, real systems do equilibrate, possibly explaining
the experimentally observed corrections to quantized conductance \cite{0.7}.
Equilibration of one-dimensional boson systems was recently studied in atomic
traps \cite{traps}.

A finite life time of excitations in the Luttinger liquid can be understood if
small anharmonic corrections are added to the Hamiltonian.  Such perturbations
are irrelevant in the sense that their effect rapidly decreases as the
temperature approaches zero.  However, they are responsible for the
interaction of bosonic excitations and therefore for their equilibration.
Scattering of the excitations caused by the anharmonic coupling terms
preserves not only their total energy but also momentum.  Thus the resulting
equilibrium distribution of the bosonic excitations
\begin{equation}
  \label{eq:Boson_distribution}
  N_q=\frac{1}{e^{\hbar(v|q|-uq)/T}-1}
\end{equation}
is controlled by two parameters, temperature $T$ and velocity $u$.  Here $q$
is the wave vector of the excitation.  

It is important to note that translation invariance of the problem ensures
conservation of the total momentum of the system, rather than that of its
elementary excitations.  This subtle distinction can be understood by
considering the expression
\begin{equation}
  \label{eq:Luttinger_Momentum}
  P=\frac{\pi\hbar N}{L} J +\sum_q \hbar q\, b_q^\dagger b_q^{}
\end{equation}
for the momentum of a Luttinger liquid \cite{haldane,haldane2}.  Here $b_q$ is
the boson annihilation operator, $N$ is the total number of particles, $L$ is
the system size.  Periodic boundary conditions require that $J$ be an even
number if the underlying physical particles are bosons, while for fermions
$J+N$ must be even.  The first term in Eq.~(\ref{eq:Luttinger_Momentum})
accounts for the momentum associated with the motion of the system as a whole,
which is possible even in the absence of excitations.

Unless additional conservation laws are present, one should expect the
existence of scattering processes which transfer momentum between the
excitations and the system as a whole.  The minimum momentum transfer $\Delta
p=2\pi\hbar N/L$ corresponds to $J$ changing by 2.  Because the typical
momentum of an excitation $\hbar q\sim T/v$ is small at $T\to 0$, such
processes involve a large number of excitations.  They are not included in the
standard Luttinger liquid theory.  Although their rate is small, these
processes are required for the full equilibration of the Luttinger liquid.
Physically one expects them to lead to relaxation of the velocity $u$ of the
gas of excitations in Eq.~(\ref{eq:Boson_distribution}) towards an equilibrium
value $v_d$,
\begin{equation}
  \label{eq:relaxation}
  \dot u = -\frac{u-v_d}{\tau}.
\end{equation}
We limit our consideration to Galillean invariant systems of particles, whose
mass is denoted by $m$.  In this case the system must be at rest in a
reference frame moving with the center of mass, and $v_d=P/mN$.  The study of
the relaxation time $\tau$ is the main goal of this paper.

\begin{figure}[t]
 \resizebox{.48\textwidth}{!}{\includegraphics{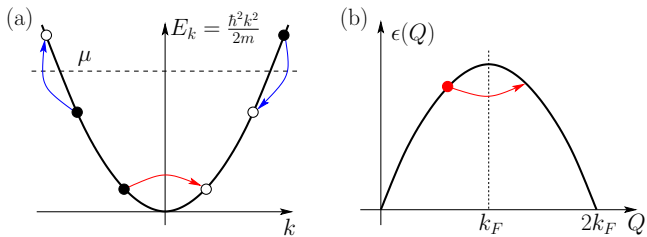}}
 \caption{\label{fig:three-particle} (a) The simplest backscattering process
   in a weakly interacting Fermi gas involves three particles, including one
   near the bottom of the band. (b) Backscattering of a fermion at the bottom
   of the band can be interpreted as a hole excitation overcoming a barrier at
   $Q=k_F$.}
\end{figure}

We start by reviewing the simplest case of a system with Luttinger liquid
behavior at low energies, namely, the weakly interacting Fermi gas.  In
fermionic Luttinger liquids the integer $J$ can be interpreted
\cite{haldane} as the difference of the numbers of right- and left-moving
fermions, $J=N^R-N^L$.  Clearly, the scattering process changing $J$ by 2
involves backscattering of a fermion, $\Delta N^R=-\Delta N^L=\pm1$.  Because
of conservation of energy and momentum, two-particle scattering in one
dimension results only in particles exchanging their momenta, and the
distribution function remains unchanged.  Thus the simplest scattering process
involves three particles, see Fig.~\ref{fig:three-particle}(a).  Simultaneous
conservation of momentum and energy requires involvement of hole states below
the Fermi level.  At $T\to0$ the most efficient process involves a hole near
the bottom of the band, whose scattering is accompanied by creation and
collapse of particle-hole pairs with energies of order $T$ near the two Fermi
points \cite{lunde}.  Since the occupation probability of a hole state near
$k=0$ is exponentially small, one finds a small equilibration rate
$\tau^{-1}\propto e^{-E_F/T}$, where $E_F=\hbar^2k_F^2/2m$ is the Fermi
energy \cite{micklitz}.

It is instructive to rephrase the above argument in the language of a hole
excitation with wave vector $Q$ and energy $\epsilon(Q)=\hbar v_FQ(1-Q/2k_F)$
constructed by moving a fermion from state $k_F-Q$ to the Fermi level state
$k_F$.  (Here $v_F$ is the Fermi velocity.)  The hole is scattered off of
particles near the Fermi level, with its momentum changing in steps of $\Delta
Q\sim T/v_F$.  Backscattering occurs when such a hole crosses the point
$Q=k_F$, Fig.~\ref{fig:three-particle}(b).

This picture can now be generalized to the case of arbitrary interaction
strength.  The particle hole pairs with momenta $\hbar q\sim T/v$ near the two
Fermi points transform into the bosonic excitations in the Luttinger liquid
\cite{haldane}.  On the other hand, the hole with the large wave vector $Q\sim
k_F$ is not accounted for by the Luttinger liquid theory and should be treated
as a mobile impurity \cite{pustilnik,imambekov}.  In the presence of
interactions its energy $\epsilon(Q)$ is defined as that of the lowest energy
state of momentum $\hbar Q$, measured from the ground state.  Throughout this
paper we assume that $\epsilon(Q)$ remains convex.  Then the equilibration
rate shows activated temperature dependence $\tau^{-1}\propto
e^{-\epsilon(k_F)/T}$, where $k_F=\pi n_0$ is determined by the average
particle density $n_0=N/L$.

To obtain a full expression for the equilibration rate, the distribution
function of the holes should be considered carefully.  To first approximation
it can be obtained by noticing that the holes are scattered by the bosonic
excitations, distributed according to Eq.~(\ref{eq:Boson_distribution}).
These scattering events involve exchange of both energy and momentum between
the hole and the bosons, leading to the equilibrium distribution
\begin{equation}
  \label{eq:boundary_conditions}
  f(Q)\simeq
  \left\{
  \begin{array}[c]{ll}
    e^{-\epsilon_u(Q)/T}, & Q<k_F,
\\
    e^{-[\epsilon_u(Q)+2\hbar k_F u]/T}, & Q>k_F,
  \end{array}
  \right.
\end{equation}
where $\epsilon_u(Q)=\epsilon(Q)-\hbar uQ$. The apparent asymmetry between the
cases of right- and left-moving holes, $Q<k_F$ and $Q>k_F$, is caused by our
convention to measure the momentum $Q$ of the hole from the right Fermi point,
$k=+k_F$.

The discontinuity of the hole distribution function
(\ref{eq:boundary_conditions}) at $Q=k_F$ originates from the implicit
assumption that the right- and left-moving holes are distinct particles.  In
reality, the backscattering processes shown in Fig.~\ref{fig:three-particle}
convert right-moving holes into left-moving ones, thereby smearing the
discontinuity of the distribution function $f(Q)$.  Because the hole moves in
momentum space via random small steps of $\Delta Q\sim T/\hbar v$, this motion
is diffusive.  Such diffusion was considered previously for the cases of
weakly-interacting \cite{micklitz} and strongly-interacting
\cite{equilibrationWigner} electrons.  It is described by the Fokker-Planck
equation
\begin{equation}
  \label{eq:Fokker-Planck}
  \partial_t f = -\partial_Q J,
\quad
  J=-\frac{B(Q)}{2}\left[\frac{\epsilon_u'(Q)}{T}+\partial_Q\right]f,
\end{equation}
where the expression for the probability current $J$ assumes that the system
as a whole is at rest, $v_d=0$, and prime denotes the derivative with respect
to $Q$.  The diffusion constant in momentum space
\begin{equation}
  \label{eq:B_definition}
  B(Q)=\sum_{\delta Q}[\delta Q]^2 W_{Q,Q+\delta Q}
\end{equation}
is defined in terms of the rate $W_{Q,Q+\delta Q}$ of scattering events
changing the wave vector of the hole from $Q$ to $Q+\delta Q$.

We now find a stationary solution of the Fokker-Planck equation with the
boundary conditions (\ref{eq:boundary_conditions}), which gives a uniform in
$Q$-space probability current
\begin{equation}
  \label{eq:J}
  J=uB(k_F)\frac{\hbar k_F}{T}
    \left(\frac{|\epsilon''(k_F)|}{2\pi T}\right)^{1/2}
    e^{-\epsilon(k_F)/T}.
\end{equation}
Here to obtain the expression for $J$ to leading order in $u$ we neglected the
difference between $\epsilon(Q)$ and $\epsilon_u(Q)$.

A non-zero probability current $J$ means that the holes backscatter at a rate
$JL/2\pi$.  With each backscattering event transferring momentum $\Delta
p=2\hbar k_F$ from excitations to the motion of the system as a whole, we find
$\dot P_{\rm ex}=-JL\hbar k_F/\pi$.  Comparing this result with the expression
$P_{\rm ex}=(\pi L T^2/3\hbar v^3)u$ for the total momentum of the excitations
obtained using the distribution (\ref{eq:Boson_distribution}), we find the
relaxation law $\dot u=-u/\tau$ with the rate
\begin{equation}
  \label{eq:equilibration_rate}
  \tau^{-1}= \frac{3\hbar k_F^2  B}{\pi^2\sqrt{2\pi m^*T}} 
            \left(\frac{\hbar v}{T}\right)^3  e^{-\Delta/T}.
\end{equation}
Here $\Delta=\epsilon(p_F)$, the effective mass of the hole
$m^*=-\hbar^2/\epsilon''(k_F)$, and the diffusion constant $B=B(k_F)$ remains to
be determined.

Following Refs.~\cite{pustilnik,imambekov}, we treat the hole in a Luttinger
liquid as a mobile impurity.  The Fokker-Planck equation for such an impurity
was discussed in Ref.~\cite{castroneto}.  The parameter
$B$ was found to scale as
\begin{equation}
  \label{eq:B_vs_T}
  B=\chi T^5
\end{equation}
at $T\to0$.  The approach of Ref.~\cite{castroneto} does not allow for the
determination of the coefficient $\chi$.  The latter is controlled by the
interactions between the physical particles forming the Luttinger liquid.  In
the limit of strong Coulomb repulsion it was calculated in
Ref.~\cite{equilibrationWigner}.  A related calculation was performed in the
context of decay of dark solitons in weakly-interacting one-dimensional Bose
systems \cite{gangardt}.

Our next goal is to obtain an exact expression for the coefficient $\chi$ in
Eq.~(\ref{eq:B_vs_T}) for arbitrary interactions between the particles forming
the Luttinger liquid.  Microscopically the case of arbitrary interaction
strength can be approached only for integrable systems, where an infinite
number of conservation laws allows one to diagonalize the Hamiltonian exactly.
However, the same conservation laws ensure that the excitations have infinite
life times and $B=0$.  We thus develop a phenomenological theory and express
$B$ in terms of hole spectrum $\epsilon(Q)$.

We describe the system in terms of the displacement $u(y)$ of a small element
of the liquid from its reference position $y$ in a state of uniform particle
density $n_0$, and the conjugate momentum density $p(y)$ such that
$[u(y),p(y')]=i\hbar\delta(y-y')$.  In the absence of the hole excitations the
Hamiltonian of the liquid can be written as
\begin{equation}
  \label{eq:H_0}
  H_L=\int\left[
         \frac{p^2}{2mn_0} + n_0U(n)
        \right] dy,
\end{equation}
where $U(n)$ is the internal energy per particle, determined by the
fluctuating density $n(y)=n_0/[1+u'(y)]$.  Expanding (\ref{eq:H_0}) up to the
third order in small deformation $u'$ one finds
\begin{equation}
  \label{eq:H_0_approximate}
  H_L=\int\left(
         \frac{p^2}{2mn_0} + \frac{mn_0v^2}{2}\,u'^2 -\alpha u'^3
        \right) dy.
\end{equation}
Here the sound velocity $v=[(2n_0U'+n_0^2U'')/m]^{1/2}$ and
$\alpha=n_0^2U'+n_0^3U''+n_0^4U'''/6$.  The quadratic part of
Eq.~(\ref{eq:H_0_approximate}) is the Hamiltonian of the Luttinger liquid,
which can be brought to the form $\sum \hbar v|q|b_q^\dagger b_q$ by
introducing the boson operators $b_q$ via the standard procedure
\begin{subequations}
\begin{eqnarray}
  \label{eq:bosons_u}
  u(y)&=&\sum_q\sqrt{\frac{\hbar}{2mn_0Lv|q|}}\,
       (b_qe^{iqy}+b_q^\dagger e^{-iqy}),
\\
  \label{eq:bosons_p}
  p(y)&=&-i\sum_q\sqrt{\frac{\hbar mn_0v|q|}{2L}}\,
       (b_qe^{iqy}-b_q^\dagger e^{-iqy}).
\end{eqnarray}
  \label{eq:bosons}
\end{subequations}

The presence of a hole excitation at the point in the liquid with reference
position $Y$ is accounted for by adding a term
$H_h=\epsilon(Q)=\epsilon(-i\partial_Y)$ to the Hamiltonian
(\ref{eq:H_0_approximate}).  Since our goal is to evaluate $B=B(k_F)$, we
assume that $Q$ is near $k_F=\pi n_0$ and use the expansion
\begin{equation}
  \label{eq:spectrum_expansion}
  H_h=\Delta(n(Y))-\frac{\hbar^2}{2m*}(-i\partial_Y-\pi n_0)^2.
\end{equation}
It is worth mentioning that our Hamiltonian is written in terms of the
Lagrangian variable $u(y)$ defined as function of reference position $y$,
rather than Eulerian variable $n(x)$ at the physical position $x=y+u(y)$.  The
two approaches are, of course, equivalent and lead to the same results
\cite{unpublished}.  Although the use of Eulerian variables is more common in
the Luttinger liquid theory, our method has the advantage of more simply
accounting for the Galilean invariance of the problem.  In addition, since $Y$
is the position of the hole in the reference state of uniform density $n_0$,
the maximum of $\epsilon(Q)$ is located at $Q=\pi n_0$, regardless of the
physical density $n$.  On the other hand, the maximum value $\Delta$ is a
function of $n=n_0/(1+u')$.  This dependence gives rise to interaction of the
hole with the Luttinger liquid.  Expanding (\ref{eq:spectrum_expansion}) to
second order in $u'$, we obtain
\begin{equation}
  \label{eq:perturbation}
  H_h=-\beta_1 u'(Y) +\beta_2[u'(Y)]^2
      -\frac{\hbar^2}{2m^*}(-i\partial_Y-\pi n_0)^2,
\end{equation}
where $\beta_1=n_0\Delta'$, $\beta_2=n_0\Delta'+n_0^2\Delta''/2$, and we
omitted the constant $\Delta(n_0)$.

In order to find the diffusion constant in momentum space $B(k_F)$,
Eq.~(\ref{eq:B_definition}), we evaluate the scattering rate $W_{Q,Q+\delta
  Q}$.  The momentum of the hole changes as it interacts with the bosonic
excitations, see Eqs.~(\ref{eq:perturbation}) and (\ref{eq:bosons}).  The
processes involving one boson cannot simultaneously conserve both energy and
momentum of the system.  The simplest allowed process for a hole near $Q=\pi
n_0$ involves absorption of a boson $q_1$ and simultaneous emission of a boson
$q_2$ such that $q_2\approx-q_1$ \cite{castroneto,gangardt,pustilnik}.  The
scattering rate is then found from the Fermi golden rule expression
\begin{eqnarray*}
  \label{eq:W}
  W_{Q,Q+\delta Q}&=&\frac{2\pi}{\hbar}
                  \sum_{q_1,q_2}|t_{q_1,q_2}|^2 N_{q_1}(N_{q_2}+1)
                  \delta_{q_1-q_2, \delta Q}
\nonumber\\
                &&\times
\delta(\epsilon(Q)-\epsilon(Q+\delta Q)+\hbar v|q_1|-\hbar v|q_2|).
\end{eqnarray*}
The matrix element $t_{q_1,q_2}$ accounts for all processes that destroy boson
$q_1$ and create boson $q_2$.  For example, a contribution proportional to
$\beta_2 b_{q_2}^\dagger b_{q_1}$ is found in the second term in
Eq. (\ref{eq:perturbation}).  Identical scattering processes can be obtained
in the second-order perturbation theory with amplitudes proportional to
$\beta_1^2$ or $\alpha\beta_1$.  The calculation is simplified considerably by
applying to the Hamiltonian the unitary transformation $U^\dagger (H_L+H_h)U$
with
\begin{equation}
  \label{eq:unitary}
  U=\exp\left(
         \frac{i\beta_1}{\hbar mn_0v^2}\int_{-\infty}^Yp(y)dy
        \right).
\end{equation}
This removes the $-\beta_1u'(Y)$ term in (\ref{eq:perturbation}) and generates a
correction to $\beta_2$ proportional to $\alpha\beta_1$.  In addition, a new
term $p^2(Y)$ is generated with the coefficient proportional to $\beta_1^2$.
Both the $[u'(Y)]^2$ and $p^2(Y)$ terms contain contributions of the form
$b_{q_2}^\dagger b_{q_1}$ and give rise to the matrix element
\begin{equation}
  \label{eq:t_q1q2}
  t_{q_1,q_2}=-\frac{\hbar\sqrt{|q_1q_2|}}{mn_0Lv}
             \left(
              \beta_2-\frac{3\alpha\beta_1}{mn_0v^2}+\frac{\beta_1^2}{2m^*v^2}
             \right).
\end{equation}
As a result, we recover the temperature dependence (\ref{eq:B_vs_T}) with the
coefficient $\chi$ given by
\begin{equation}
  \label{eq:chi}
  \chi=\frac{4\pi^3n_0^2}{15\hbar^5m^2v^8}
       \left(
        \Delta''-\frac{2v'}{v}\Delta'+\frac{\Delta'^2}{m^*v^2}
       \right)^2,
\end{equation}
where prime denotes the derivative with respect to the particle density $n_0$.

The above result completes our evaluation of the relaxation rate of a
Luttinger liquid, given by Eqs. (\ref{eq:equilibration_rate}),
(\ref{eq:B_vs_T}), and (\ref{eq:chi}).  The rate has activated temperature
dependence with both the activation temperature $\Delta$ and the prefactor
determined by the spectrum of holes $\epsilon(Q)$.  Although our result is
applicable at any interaction strength, the spectrum $\epsilon(Q)$ is known
analytically only in a few special cases.  

For non-interacting spinless fermions $\Delta$ is given by the Fermi energy
$(\pi\hbar n_0)^2/2m$, $v$ is the Fermi velocity $\pi\hbar n_0/m$, and
$m^*=m$.  This results in $\chi=0$, as there is no scattering of holes in the
absence of interactions.  In the limit of weak interactions, the spectrum
$\epsilon(Q)$ should be evaluated up to second order in interaction strength.
This gives rise to a result \cite{unpublished} for $\chi$ consistent with the
rather complicated expression for the three-particle scattering amplitude
\cite{lunde} that controls the scattering of holes,
Fig.~\ref{fig:three-particle}(a).  In the limit of strong long-range
repulsion, the system forms a Wigner crystal, and the hole spectrum coincides
with that of phonons in the crystal.  We have verified that in this regime our
expression (\ref{eq:chi}) recovers the results of
Ref.~\cite{equilibrationWigner}.  We have also found that in the case of
weakly interacting bosons Eq.~(\ref{eq:chi}) is consistent with the expression
for the mobility of the so-called dark soliton \cite{gangardt}.

At arbitrary interaction strength the spectrum of holes is known only for
integrable models.  As we already mentioned, integrability means absence of
scattering of excitations, $B=0$.  We have verified that our expression
(\ref{eq:chi}) vanishes for the Calogero-Sutherland model of particles with
inverse-square repulsion \cite{sutherland}, and for the Lieb-Liniger model of
bosons with point-like repulsion \cite{liebliniger}.

The authors are grateful to M. Pustilnik for discussions.  This work was
supported by the U.S. Department of Energy under Contract
Nos. DE-AC02-06CH11357 and DE-FG02-07ER46452.

\end{document}